\documentclass[a4paper,fleqn,usenatbib]{mnras}

\usepackage{newtxtext,newtxmath}

\usepackage{etoolbox}

\usepackage{xcolor}
\usepackage{amsmath}
\usepackage{graphicx}
\usepackage{psfrag}

\newcommand{\beq}{\begin{equation}}
\newcommand{\eeq}{\end{equation}}
\newcommand{\nn}{\nonumber}

\newcommand{\brac}[1]{\left({#1}\right)}
\newcommand{\pd}[2]{\frac{\partial{#1}}{\partial{#2}}}

\newcommand{\curl}{\nabla\times}

\newcommand{\br}{{\boldsymbol r}}
\newcommand{\bB}{{\boldsymbol B}}

\newcommand{\bv}{{\boldsymbol v}}

\newcommand{\Pm}{\textrm{Pr}_\textrm{m}}
\newcommand{\Rm}{\textrm{Rm}}
\newcommand{\bOm}{{\boldsymbol{\Omega}}}
\newcommand{\dbxi}{\dot{\boldsymbol{\xi}}}

\newcommand{\skl}[1]{{\color{black}{#1}}}

\title[Three NS dynamo phases]{Generating neutron-star magnetic
  fields: three dynamo phases}
\author[S. K. Lander]{S. K. Lander\thanks{samuel.lander@uea.ac.uk}\\ \\
         School of Physics, University of East Anglia, Norwich, NR4 7TJ, U.K.}

\begin{document}

\pagerange{\pageref{firstpage}--\pageref{lastpage}} \pubyear{0000}
\maketitle

\label{firstpage}

\begin{abstract}
Young neutron stars (NSs) have magnetic fields in the range
$10^{12}-10^{15}$ G, believed to be generated by dynamo
action at birth. We argue that such a dynamo is actually too
inefficient to explain the strongest of these fields. Dynamo
action in the mature star
is also unlikely. Instead we propose a promising new precession-driven
dynamo and examine its basic properties, as well as arguing
for a revised mean-field approach to NS dynamos. The
precession-driven dynamo could also play a role in field generation in main-sequence stars.
\end{abstract}

\begin{keywords}
dynamo -- stars: evolution -- stars: magnetic fields -- stars: neutron -- stars: rotation
\end{keywords}

\maketitle


\section{Introduction}

The strongest long-lived magnetic fields $B$ in the Universe are hosted by
neutron stars (NSs), with inferred external dipole fields
$B_{\rm dip}$ up to $10^{15}$ G in strength, but usually considerably weaker.
For older NSs some of this variability can be attributed to accretion
or field evolution, but even among young NSs the variation 
is huge: for example, the Crab pulsar and the magnetar SGR 1806-20 are both around 1000 yr old,
but the former has $B_{\rm dip}=3.8\times 10^{12}$ G and the latter
$B_{\rm dip}\approx(0.8-2)\times 10^{15}$ G.
The physics that leads to some NSs -- but not others -- having $B_{\rm dip}\sim 10^{15}$ G
is still poorly understood, despite being of interest both on theoretical
grounds, and for observed astrophysical
phenomena. For example, early generation of a strong $B$ is
essential for leading models of superluminous supernovae,
$\gamma$-ray bursts and afterglows \citep{zhang_mesz,thomp04}, 
and might result in the star emitting detectable gravitational
radiation \citep{stella05}.

The simplest explanation for NS magnetism is that it is a `fossil',
inherited from the degenerate iron core of its progenitor
star. \skl{Magnetic flux scales with the square of the stellar radius $R_*$;
  assuming this is conserved in the compression of a typical
  progenitor ($R_*\sim 10^6$ km) to NS proportions
  ($R_*\sim 10$ km) yields a factor-$10^{10}$
  amplification of $B$. One can thus account for
  the highest $B$ in NSs, since there do indeed exist main-sequence stars with $B$ up to $\sim 10^4$ G
  \citep{land92,donati_land} -- although the actual amplification
  factor will be less than $10^{10}$, since only the flux of the
  progenitor's iron core is inherited by
  the NS. Immediately before collapse this core has a radius
  $\sim 3000$ km \citep{sukhbold16}, so within the fossil field
  scenario its $B$ would in some cases be as high as $(10/3000)^2\times
  10^{15}\textrm{ G}=10^{10}\textrm{ G}$. NS magnetic fields are
  unlikely to be fossils, however, for at least two reasons: }
 there appear to be too few suitable progenitors
to explain the likely magnetar birth rate \citep{makarenko}; 
and the violent dynamics in which a NS
is born will not leave $B$ unaffected.

If $B$ is not a fossil,
then, it must be amplified after birth to the kind of strengths we
see. Large-scale efficient amplification of $B$ requires a
\emph{dynamo}: a mechanism for converting kinetic energy of (generally)
small-scale fluid motions to magnetic
energy \citep{moffatt78,rincon}.
NSs have
extremely low electrical resistivity $\eta$, and so field lines are
`frozen' into the fluid. A dynamo exploits this: any field lines
threading a fluid element are stretched as the fluid moves, thus
increasing the magnetic energy. At the same time, this amplified and
distorted field needs $\eta\neq 0$ in order to reconnect
in its new geometry. We will show that reconnection is
particularly difficult in the unique physical conditions
of a NS.

Here we assess the possibility of dynamo action in a NS during three
different phases of its life, with a view 
to understand how and when its intense $B$ is generated.
We argue for a revision of dynamo theory for this problem, show how
the usual scenario for magnetic-field generation in NSs may not work,
and present a promising new alternative.

\section{Phase 1: the hot convective star}

In its very early life a NS experiences differential
\skl{rotation (e.g. \citet{janka_moench}), a consequence of 
  the approximate conservation of angular momentum of cylindrical
  shells of matter during core collapse.}
In addition, its outer half will be convectively unstable
\citep{epstein79}. These are just the ingredients needed for dynamo
action, with the prospect of amplifying the star's large-scale $B$ up to $\sim 10^{15}$ G
by drawing on kinetic energy from turbulent
motions \citep{TD93}.

\skl{A stellar dynamo involves dynamics over a
  wide range of lengthscales, from $R_*$ down to a typically
  microscopic scale related to e.g. damping/reconnection, making it
  intractable to study without some kind of approximation. One either 
  attempts a high-resolution direct numerical simulation, hoping that the
  unresolvable fine lengthscales are not essential for the dynamo, or
  uses lower resolution with a method to explicitly account 
  for the effect of the subgrid physics. Studies of magnetic-field
  generation often employ mean-field
theory \citep{krause_raed}, an averaging procedure in which $\bB$ and
the velocity $\bv$ are split into large-scale mean-averaged
quantities, and small-scale fluctuating/stochastic terms.} Only the
former (here denoted with overbars) directly enter the field-evolution equations, with the
influence of the latter felt
through a \emph{dynamo closure} relation -- essentially an assumption
about the form the small-scale $\bv\times\bB$ term takes upon
mean-averaging. In particular, turbulent convection is accounted for
through an additional `$\alpha$ effect' term in the mean-$B$ evolution:
\beq
\pd{\bar\bB}{t}=\curl\left[\bar\bv\times\bar\bB+\alpha\bar\bB-\eta\curl\bar\bB\right]
\eeq  
(a similar result may be derived
in general relativity; \citet{buccdelz13}). $B$ can be amplified through the joint action
of turbulent convection with differential rotation -- an `$\alpha-\Omega$
dynamo' -- or by convection alone through an `$\alpha^2$
dynamo'. \skl{Evolving the mean-field equations, \citet{bonrezurp}
  find that} the dominant dynamo effect for a proto-NS seems to vary
with rotation rate. \skl{Recent direct numerical simulations give
  additional information: that} although the highest $B$ are generated in
rapidly-rotating models \citep{raynaud}, some dynamo activity is still
present at slower rotation \citep{masada20}.

\skl{Two dimensionless numbers} are key to understanding NS dynamos: the
magnetic Reynolds and magnetic Prandtl numbers, $\Rm$
and $\Pm$ respectively. $\Rm\equiv v_{\rm char}l_{\rm char}/\eta$ gives the ratio of advection to
diffusion of $B$ by the flow
(where $v_{\rm char}$ and $l_{\rm char}$ are the characteristic
velocity and lengthscale of the flow),
and $\Pm\equiv \nu_{\rm s}/\eta$ shows the relative importance of \skl{kinematic shear}
viscosity $\nu_{\rm s}$ to resistivity in dissipating energy of the magnetised
fluid. \skl{In a
  young NS\footnote{Throughout this paper, unless stated otherwise,
    we report typical numerical values for a proto-NS core, with
    $\rho\approx 10^{14}\textrm{g cm}^{-3},T\approx 10^{10}-10^{11}\textrm{ K}$.}
  $\eta$ is primarily due to electron-proton scattering, with a typical value of
  $10^{-6}-10^{-4}\textrm{ cm}^2\textrm{s}^{-1}$
  \citep{BPP69b,raynaud}. The main contribution to $\nu_{\rm s}$ changes
  depending on whether the stellar matter has cooled enough to be
  neutrino-transparent, a transition that occurs within a minute from birth. If so, neutron-neutron
  scattering dominates and $\approx 1\textrm{ cm}^2\textrm{s}^{-1}$
  \citep{cutlind}; if not, neutrino-nucleon scattering dominates,
  leading to a much higher $\nu_{\rm s}\approx 10^8\textrm{ cm}^2\textrm{s}^{-1}$
  \citep{keil96}. The latter, neutrino-opaque, regime is relevant for
  a proto-NS, and so $\Pm\sim 10^{13}$. To find $\Rm$ we follow \citet{TD93} and take
  $v_{\rm char}=10^8\textrm{cm s}^{-1},l_{\rm char}=10^5\textrm{cm}$,
  yielding $\Rm\sim 10^{17}$.}

The best-understood dynamos are `slow', with 
growth rates \skl{that tend to zero as
$\Rm\to\infty$. `Fast' dynamos, by constrast, still generate $B$ in this
limit, even if they cannot be truly non-diffusive
\citep{moffproc}; the archetypical example is the 
stretch-twist-fold dynamo \citep{vainzel}. Rigorous analysis
is difficult, but any dynamo in a NS must -- given their enormous
$\Rm$ -- be fast, so we will assume that results for both
fast and high-$\Rm$ dynamos (in principle distinct notions) are relevant here.}

Any dynamo has to create magnetic flux more quickly than it is dissipated,
suggesting that a large $\Rm$ is helpful -- but the huge values
associated with proto NSs in particular
are, in fact, problematic. At least some such 
dynamos involve chaotic fluid motions that result in fractally-distributed
$B$ with a strongly fluctuating
direction \citep{finnott}; reconnection could then cause local cancellations of parallel and
antiparallel field vectors, leaving a weak large-scale $B$. Indeed, 
\cite{vaincatt} found that high-$\Rm$ dynamos saturate at
values of flux too low to explain typical astrophysical $B$.
\skl{A related concern is how a
  large-scale $B$ can be rearranged, given that the low $\eta$
  suggests a microscopic reconnection scale. This was allayed by \citet{LV99}, who showed that
  high-$\Rm$ MHD turbulence with a weak stochastic component 
  does allow for fast reconnection of the large-scale $B$ -- and
  \citet{parker92} argues that fast reconnection in turn supports a fast dynamo.}
Furthermore, an inverse cascade effect can convert
small-scale helicity into large-scale $B$ \citep{frisch,brand01}.
Together, these studies give confidence in the ability of high-$\Rm$ dynamos
to amplify large-scale $B$, and also suggest that numerical
simulations of astrophysical dynamos -- which necessarily employ
unphysically small $\Rm$ -- are nonetheless faithful to the
astrophysical phenomena they intend to represent.

Key to these results, however, is that $\Pm$ is small, as is
the case for non-degenerate stars but emphatically not for NSs. At
large $\Pm$ turbulence will tend to be viscously smoothed
out on lengthscales longer than those on which reconnection takes
place. This causes a reduction in reconnection speed by a
factor \citep{jafari18}
\beq\label{Pm_reduce}
\frac{\Pm^{-1/2}}{1+\ln(\Pm)}
\eeq
compared with the $\Pm=1$ case; for a proto-NS the reduction factor is
$10^8$, \skl{and the effect on dynamo action may be similarly
  deleterious.}
In addition, at large $\Pm$ the inverse cascade effect is replaced by a
`reversed dynamo', in which conversion of magnetic to kinetic energy
occurs at short lengthscales \skl{ \citep{brand_remp}, potentially}
thwarting \skl{efficient} large-scale field amplification.

Pessimistically, one could therefore envisage that whilst a real proto-NS
dynamo amplifies a small-scale multidirectional $B$, this is then substantially
annulled as it slowly reconnects, never managing to amplify the
large-scale $B$.
Furthermore, the work of \cite{jafari18} and \cite{brand_remp}
suggests that typical proto-NS dynamo simulations -- in which $\Rm$, $\Pm$ are factors
of $\sim 10^{16},10^{11}$ (respectively) too small -- may not be
representative of the real system\footnote{Some such
  simulations \skl{(e.g. \citet{moesta})} evolve the ideal MHD
  ($\eta=0$) equations, relying on the unphysical
artefact of numerical resistivity to provide reconnection. These
`ideal' simulations therefore have dissipation on the grid
spacing ($10^3+$ cm) and so an effective $\Rm\lll 10^{17}$.}.

These issues would vex not only the convective dynamo, but also
any other field-amplification 
mechanism during this phase: 
e.g. one driven by the magneto-rotational instability
\citep{ober09,sawai13,moesta,reboul21}
or the Tayler-Spruit dynamo \citep{spruit02}.

How can we understand the details of a proto-NS dynamo, if realistic
$\Rm$ and $\Pm$ values are unattainable in a numerical approach? One
possibility could be evolutions employing a revised mean-field dynamo that reflects the unique small-scale
conditions of a high-$\Rm$, high-$\Pm$ dynamo through a suitable closure
relation. It is known that such conditions tend to produce a field
concentrated into flux ropes \citep{gallprocweiss}, which is analogous
to a similar problem in the context
of a mature NS core, where type-II superconductivity quantises the
local field into thin fluxtubes. The global magnetic-field evolution
of this latter problem has been studied in some detail
\citep{mendell98,graber}, and provides a promising starting point for 
revising the NS dynamo equations.


\section{Phase 2: the warm precessing star}

Precession was originally proposed as a possible mechanism for driving the
geodynamo \citep{bullard49,malkus68}, and both numerical models
\citep{tilgner05,tilgner07,wu_roberts} and laboratory experiments
\citep{giesecke18} have established its viability for amplifying
$B$; in all cases a solid boundary precesses and drives internal
  fluid motion.
Precession consists of a vector sum of rotations about two axes:
\beq
\bOm=\bOm_0+\bOm_{\rm p}
\eeq
where $\bOm_0,\bOm_{\rm p}$ are the primary and secondary rotations. In
literature on fluid dynamics, the ratio of these two is often called
the Poincar\'e number $\textrm{Po}=\Omega_{\rm p}/\Omega_0$, with a typical
value being $\textrm{Po}=0.1$.

A NS can undergo free precession (i.e. no external driving force)
due to the presence of a distortion misaligned from $\bOm_0$
by some angle $\chi$. Often this is assumed to be an elastic asymmetry
in the star's solid crust \citep{jonesand01}, but by the time the star has
cooled enough for this to form, dynamo action may well be totally
suppressed, as discussed later.

Here we describe a new precession-driven dynamo that can operate
  in an entirely fluid body, applying the idea to a young NS. It uses
  the key result that the star's $B$ always
  induces some distortion (or `rigidity') $\epsilon_B\propto B^2$ that is typically misaligned
  from $\bOm_0$ by some angle $\chi$ and thus drives precession \citep{spitzer}.

A dominantly poloidal (toroidal) field induces an oblate (prolate)
distortion. The two cases have
different minimum-energy states: $\chi=0^\circ$ ($90^\circ$) for an
oblate (prolate) body. Now, once the proto-NS phase has
finished it is likely that differential rotation will have wound up
the birth $B$ to leave a strong toroidal component $B_{\rm tor}$ roughly
symmetric about $\bOm_0$ (i.e. $\chi\approx 0^\circ$ afterwards).
We will therefore
regard this $B_{\rm tor}$ component as dominant, so that the star has a tendency for
$\chi$ to \emph{increase} towards its minimum-energy state of
$\chi=90^\circ$, and so to precess spontaneously. Purely toroidal
fields are, however, unstable \citep{tayler73} -- and so we assume the
presence of a poloidal component $B_{\rm pol}$ weak enough to 
be neglected in the first instance, but strong enough to stabilise the overall $B$.
Using a solution for $\epsilon_B$ of a toroidal field \citep{LJ09}, we may then
calculate:
\beq
{\rm Po}=\frac{\Omega_{\rm p}}{\Omega_0}=\frac{\Omega_0|\epsilon_B|\cos\chi}{\Omega_0}
 = 3\times 10^{-6}\brac{\frac{B_{\rm tor}}{10^{15}\ \textrm{G}}}^2\cos\chi,
\eeq
clearly far smaller than in the fluid-dynamics context.

Understanding how long the precession phase lasts requires a more
detailed look at the dynamics of a young magnetised NS. Although its
bulk motion is precession, within the star this must be 
supported by a complicated field of
hydromagnetic motions $\dbxi$ \citep{mestel_takh}, with the first
self-consistent solution being found by \cite{LJ17}. Secular
viscous damping of 
$\dbxi$ reduces the precessional kinetic energy, and thus causes
the evolution of $\chi$ towards $90^\circ$ for our assumed
dominantly-toroidal $B$ \citep{jones75}. Solutions of the coupled
$\Omega-\chi$ differential equations indicate that the phase of
increasing $\chi$ happens around 100 s after birth, when the temperature
$T\sim 10^{10}$ K \citep{LJ20}.

Precession alone can amplify both components of $B$, \skl{but since we
  anticipate that $B_{\rm tor}$ will already be large, we are most interested in how much
  $B_{\rm pol}$ (potentially considerably weaker) can
  catch up. $B_{\rm pol}$ is also the field component that extends beyond the
  star, connecting to the surface dipole value $B_{\rm dip}$ we
  estimate from NS spindown, and whose factor-1000} range of strengths
we wish to explain. The convection-like structure of 
$\dbxi$ (see fig. 8 from \cite{LJ17}) 
is already promising for dynamo action: simulations
of fully-convective M stars show that the interplay 
of (uniform) rotation with relatively slow convection can
lead to a strong large-scale axisymmetric $B$ \citep{browning08}.

To understand the effect of possible precession-driven
dynamo action on $B_{\rm pol}$, we imagine taking the stellar model of
\cite{LJ17} (precessing, with a toroidal background field) and adding
a seed poloidal field $\bB_{\rm seed}$,
which will be passively advected by the
fluid motion (on large scales given by $\dbxi$, and on small scales
probably turbulent, given the large Reynolds number
$v_{\rm char}l_{\rm char}/\nu_{\rm s} \sim 10^8$). $B_{\rm pol}$ thus undergoes a kind
of forced precession analogous to the set-up in previous work on
precession-driven dynamos. This is a reasonable first
approximation as long as $B_{\rm pol}$ is small enough for the overall
magnetic distortion to remain prolate, and for the effect on
$\dbxi$ to be negligible.

As a first step towards understanding this dynamo scenario, we
  will take the standard approach of considering its initial \emph{kinematic} phase, where
one can assume that a turbulent $\bv$ drives magnetic-field
amplification, but without considering the Lorentz force associated
with this newly-created $B$. The
small-scale turbulent $\bv$ averages to the fluid precession
solution discussed above, $\bar\bv=\dbxi$, and in the kinematic limit the induction equation becomes:
\beq
\pd{\bar\bB}{t}=\curl\left[\dbxi\times\bar\bB-\eta\curl\bar\bB\right].
\eeq
We plug into this equation an ansatz of
an exponentially-growing mode,
$\bB_{\rm pol}(\br,t)=\bB_{\rm seed}(\br){\rm e}^{t/\tau_{\rm amp}}$,
where $\tau_{\rm amp}$ the field amplification
timescale. This yields:
\beq\label{kindyn}
\frac{1}{\tau_{\rm amp}}\bB_{\rm seed}=\curl(\dbxi\times\bB_{\rm seed})
                                                                  -\curl(\eta\curl\bB_{\rm seed})
\eeq
-- an eigenvalue problem for $1/\tau_{\rm amp}$, which we assume
admits solutions with positive real part, corresponding
to exponential (dynamo) growth of $B_{\rm pol}$. In such analysis one
generally finds that dynamo action is only possible
above a certain Rm, but $\eta$ is so small for NS matter that this
will not be a limiting factor, and in this kinematic phase may be
neglected. Now rearranging eq. \eqref{kindyn} and using scalings from
\cite{LJ17}, we find that:
\beq\label{tau_amp}
\tau_{\rm amp}
   = \frac{B_{\rm seed}^2}{\bB_{\rm seed}\cdot[\curl(\dbxi\times\bB_{\rm seed})]}
 \sim \frac{l_{\rm char}}{\nu R_*\epsilon_\Omega\epsilon_B\cos\chi},
\eeq
where $\nu=\Omega/2\pi$ is the rotation rate in Hz and
$\epsilon_\Omega$ the centrifugal distortion.
For a given seed field we can quantitatively calculate
$\tau_{\rm amp}$, since we also know $\dbxi$ \citep{LJ17}.
$\bB_{\rm seed}$ is probably highly model-dependent, however, so
to maintain generality we will instead use the above
approximation in terms of $l_{\rm char}$.

We need $\tau_{\rm amp}$ to be short compared with the
duration $\tau_\chi$ of the precession phase, which in our scenario is
set by damping of $\dbxi$ due to bulk viscosity. This effect is sensitive to
$\Omega,B$ and $T$, but once $T\lesssim 10^{10}$ K the limiting
case given by eq. 61 of \cite{LJ18} becomes increasingly
accurate. Using this result and eq. \eqref{tau_amp}, we arrive
at the following criterion for significant dynamo action:
\begin{align}
1\lesssim\frac{\tau_\chi}{\tau_{\rm amp}}
 \approx  20 &\brac{\frac {1\,\textrm{cm}}{l_{\rm char}}}
                  \brac{\frac{\nu}{100\,\textrm{Hz}}}
                  \brac{\frac {10^{10}\,\textrm{K}}{T}}^6
                  \brac{\frac{B_{\rm tor}}{10^{15}\,\textrm{G}}}^4 \nn\\
                    &\times  \sin^2\chi\cos\chi.
\label{ratio}
\end{align}
We first note that if other quantities are close to the fiducial
values we use, $l_{\rm char}\lesssim 20$ cm is required for an effective
dynamo. Although the ratio increases rapidly for cooler stellar
models, dynamo action will be stifled for $T\lesssim 10^9$ K 
when the core becomes superconducting; see next section.
It is noteworthy that the dynamo depends only linearly on
rotation, but strongly on $B_{\rm tor}$ -- which suggests
that slight variations in (e.g.) the birth differential rotation could
manifest themselves as the kind of factor-1000 differences we infer in
$B_{\rm dip}$.

\begin{figure}
\begin{center}
\begin{minipage}[c]{0.8\linewidth}
  \includegraphics[width=\textwidth]{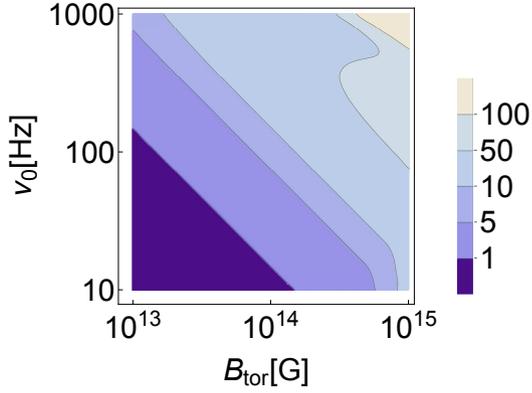}
\end{minipage}
\caption{The maximum lengthscale (colourscale; in cm) on which the precession-driven dynamo is
  effective, for a parameter space of models with
  different $B_{\rm tor},\nu_0$ as shown.
  \label{lchar}}
\end{center}
\end{figure}

As a more quantitative complement to eq. \eqref{ratio}, let us
  demand $\tau_\chi=\tau_{\rm amp}$ and rearrange eq. \eqref{tau_amp}
  to give an expression for the threshold $l_{\rm char}=l_{\rm char}^*$ for an
  effective dynamo, as a function of $\tau_\chi,\Omega,\chi$. From the
  results of self-consistent time-evolutions of the coupled $\Omega-\chi$
  equations \citep{LJ20} we find $\tau_\chi$ and $\Omega,\chi$
  time-averaged over the $\chi$-evolution phase, for models
  with different $B_{\rm tor}$ and birth rotational
  frequency $\nu_0$ \footnote{Using $B_{\rm dip}=0.01B_{\rm tor}$,
    alignment torque prefactor $k=2$, and defining $\tau_\chi$ as the era
    for which $2^\circ<\chi<88^\circ$.}. From these quantities we calculate
  $l^*_{\rm char}$, plotting the results in fig.
  \ref{lchar}. For a wide range of models we find
  $l^*_{\rm char}=1-50$ cm, in broad agreement with eq. \eqref{ratio}. Generally
  speaking $l_{\rm char}^*$ is proportional to 
  $\nu_0$ and $B_{\rm tor}$, as seen from the diagonal contours, but for
  $B_{\rm tor}> 10^{14}$ G the relationship is more complex:
  $l_{\rm char}^*$ is larger than expected for low $\nu_0$, and
  smaller than expected for high $\nu_0$. This is a manifestation of the
  nontrivial behaviour of viscous damping during
  the precession phase \citep{dallo17,LJ18}.

Dynamo action could end either through the usual mechanism of saturation
-- a backreaction of the newly-created $B$ on the flow
generating it -- or by precession ceasing, as reflected
in the trigonometric dependences in $\tau_\chi/\tau_{\rm amp}$. The
latter scenario will occur for $\chi\to 90^\circ$ 
(if $B_{\rm tor}$ remains dominant) or $\chi\to 0^\circ$ (if the newly-created
$B_{\rm pol}$ grows to become dominant, or if the star's external
alignment torque is enhanced, e.g. by fallback matter). The most optimistic
scenario would be for $B_{\rm pol}$ to grow large enough to arrest the
evolution of $\chi$ at some intermediate angle, prolonging the 
precession phase and therefore the dynamo. Since this process
ultimately taps rotational energy $E_{\rm rot}$, a firm upper limit
$B_{\rm max}$ for the increase
in average $B$-field strength is given by equating its associated
energy with $E_{\rm rot}$ and rearranging:
\beq\label{Bmax}
B_{\rm max}=\sqrt{\frac{8\pi E_{\rm rot}}{4\pi R_*^3/3}}
\approx 4\times 10^{15} \brac{\frac{\nu}{100\ \textrm{Hz}}}\ \textrm{G}.
\eeq
\skl{Let us assume, as before, that the dynamo acts
  to amplify $B_{\rm dip}$, since $B_{\rm tor}$ is already large. Then
  the above estimate suggests that it is plausible to 
  amplify a very weak $B_{\rm dip}$ to magnetar strength, but reaching
  the value $B_{\rm max}$ would need ideal conditions. If
  the dynamo is ineffective, the resulting star would still have a
  strong $B_{\rm tor}$ but a more typical pulsar-like
  $B_{\rm dip}\sim 10^{12}$ G.}
We argued in the last section that a birth dynamo may be inhibited at
high $\Pm$. Why should this precession-driven dynamo fare any better?
Firstly, since
$v_{\rm char}=\dot\xi\sim 10-10^3\textrm{ cm\ s}^{-1}$ is not so high \citep{LJ18},
$\Rm$ is a comparatively modest $\approx 10^{12}$ for the precession
phase. Perhaps more importantly though,
\beq
\Pm=2\times 10^5 (T/10^{10}\textrm{ K})^{-4}\approx 10^4-10^6
\eeq
in this case (combining results from \cite{BPP69b} and
\cite{cutlind}). Note that we calculate $\Pm$ using shear viscosity
and not the far
stronger bulk viscosity; typically it is shearing rather than compressional
motions that drive dynamo action.
The above $\Pm$ will still substantially slow down
reconnection, but by a factor $10^3$ (using eq. \eqref{Pm_reduce}) rather than the proto-NS's
$10^8$. Furthermore, since the precession phase typically lasts a factor
$100+$ longer than the convective phase\footnote{Using the code from \cite{LJ20} for
typical magnetar parameters; for extremely high $ \Omega$ the precession
phase is shortened.}, there is also less urgency
for reconnection and an inverse cascade to amplify $B$.


\section{Phase 3: the cold superconducting star}

Very little work has considered the possibility of magnetic-field
generation in the core of a mature NS, mainly because it
seems unlikely the star undergoes the kind of fluid motion needed for
a dynamo -- it is, for example, not convectively unstable. Differential rotation might, however,
persist into this late phase \citep{melatos12}, and if $\chi$ is not very close to $0^\circ$ or
$90^\circ$ precession is also possible.

If suitable fluid motions exist, the main obstacle to late-stage
dynamo action is superconductivity of the core's protons. The critical
temperature for NS superconductivity is density-dependent and poorly
constrained, but generally in the range $(1-6)\times 10^9$ K. At an
age of roughly a month to a year, most of a NS's core
will have cooled sufficiently to be
superconducting \citep{ho15}.

Intrinsic to dynamo action is that on small enough scales
magnetic-field lines must reconnect. In contrast with the case of
normally-conducting matter, the field lines in the type-II
superconducting NS core are associated with distinct physical
structures: fluxtubes. In their equilibrium state these form an
Abrikosov lattice with spacing of
$3.5\times 10^{-10}(B/10^{12}\ \textrm{G})^{-1/2}\ \textrm{cm}$ \citep{mendell98}.
Although our understanding of the physics of $B$ in the
superconducting core is still rudimentary, the dissipation (and
therefore reconnection) timescale is expected to be substantially
longer than in the normally-conducting state \citep{BPP69b}. The energy penalty for breaking a
fluxtube and the distinct inter-fluxtube spacing both hinder reconnection, which we
believe will only happen at the crust-core boundary, where
superconductivity ceases. As a result, dynamo action \emph{within}
the core seems unlikely. A dim possibility remains, however, that
fluid motions could act to bunch up fluxtubes enough for
superconductivity to be destroyed locally, thus allowing for a dynamo
in some region of limited size.

\section{Outlook}

We have argued that \skl{powerful amplification of a NS's} large-scale
$B$ is difficult during both its birth and mature
phases. Although the conditions in a proto-NS superficially resemble
those of a classic dynamo, the high values of $\Rm$ and $\Pm$ may
lead to qualitatively different -- and ineffectual -- 
action. \skl{If so, proto-NSs would never attain 
magnetar-strength $B_{\rm dip}$, casting doubt on the viability of
various models for e.g. $\gamma$-ray bursts and their afterglow
light curves.}

Magnetic-field amplification is required at some stage, however, and
this paper introduces a potentially promising new mechanism for
doing so: a precession-driven dynamo acting
$\sim 100$ s after birth. This could bypass some of the problems
associated with the proto-NS phase, may be a universal feature of a
NS's early evolution, and can naturally explain the observed
large variation of $B_{\rm dip}$ in young NSs. 
If $\chi$ evolution stalls once the dynamo stops, there is an
intriguing possibility of inferring a NS's internal magnetic-field
geometry (whether it is dominantly poloidal or toroidal) from
measurements of its present-day $\chi$.

\skl{The ideas outlined here are, however, clearly preliminary. They
  could become considerably more
  plausible through work on two key issues: the hydromagnetic dynamics
  of the precession phase, and the development of $B$ during and beyond the kinematic
  phase of the dynamo. Both of these will require numerical simulations.}

There are a number of hints from observations that a magnetar's
interior field may be considerably stronger
than its external one: \skl{\citet{makishima21} argue that long-term modulation in the
  pulse profile of a few magnetars can
  be explained by precession, if
  $B_{\rm tor}\approx 10^{16}\textrm{ G}\approx 100B_{\rm dip}$, and}
\cite{granot17} suggests that $B_{\rm tor}\gtrsim 30B_{\rm dip}$
for the magnetar Swift J1834.9-0846. 
\skl{If $B_{\rm tor}$ is so high, equation \eqref{ratio} suggests
  that amplification of $B_{\rm dip}$ during a precession-driven dynamo should
  have been relatively efficient, but the observed $B_{\rm dip}\approx
  10^{14}$ G are rather lower than expected values of $B_{\rm max}$
  from equation \eqref{Bmax}. One possibility is that the dynamo
  saturates for a poloidal field of $\sim 10^{14}$ G.}

The puzzling nature of the central compact objects, very young NSs
with $B_{\rm dip}\sim 10^{10}$ K \citep{halpgott}, could be
interpreted as the result of a failed precession-driven dynamo. Even
if $B_{\rm tor}=10^{15}$ G, say, $\chi$ could be
kept small by a strong torque due to fallback matter,
thus limiting any precession-driven amplification of $B_{\rm dip}$; if
so, the $\chi$ of these objects should still be small today.

Previous work on precession-driven dynamos has focussed on a fluid
coupled to a precessing container; in the context of the Earth, its
crust. By contrast, we have argued that a similar effect could act in
a magnetised fluid star, and therefore many of the ideas presented
here could also be viable for explaining long-term field
(re)generation in main-sequence oblique rotators.


\section*{Acknowledgements}

I thank Jonathan Granot for interesting correspondence on some of
these ideas, and the referee for a very detailed and insightful report that
substantially improved this paper.

\section*{Data availability}

The specific data underlying this article, and additional data for
other related models, will be made available upon reasonable request.

\bibliographystyle{mnras}

\bibliography{references}

\label{lastpage}

\end{document}